\documentstyle[aps,12pt,manuscript]{revtex}
\begin{document}
\preprint{INJE--TP--95--4}
\def\overlay#1#2{\setbox0=\hbox{#1}\setbox1=\hbox to \wd0{\hss #2\hss}#1%
\hskip -2\wd0\copy1}

\title{Hawking temperature from scattering off the charged 2D black hole}

\author{Jin Young Kim}
\address{Division of Basic Science, Dongseo University, Pusan 616-010, Korea}

\author{ H. W. Lee and Y. S. Myung }
\address{Department of Physics, Inje University, Kimhae 621-749, Korea}

\maketitle

\vskip 1.5in

\begin{abstract}
The charged 2D black hole is visualized as presenting an potential barrier
$V^{OUT}(r^*)$
to on-coming tachyon wave. Since this takes the complicated form, an
approximate form $V^{APP}(r^*)$
is used for scattering analysis.
We calculate the  reflection and transmission coefficients for scattering
of tachyon
 off the charged 2D black hole.
The Hawking temperature is also derived from the  reflection coefficient by
Bogoliubov transformation.  In the limit of $Q \to 0$,
we recover the
Hawking temperature of the 2D dilaton black hole.

\end{abstract}
{}~~

\newpage

Lower dimensional theories of gravity provide  the simplified contexts in
which to study
black hole physics [1].
The non-triviality of these models
arises from the non-minimal coupling of the dilaton to the scalar curvature.
A dilaton potential of the type produced by the string loop corrections may
induce multiple
horizons [2].   For example, the event (outer) horizon ($r_{+}$) as well as
inner (Cauchy) horizon
($r_{-} $) exist in the 2D charged black hole from heterotic string
theories. In this sense
the charged 2D black hole is very similiar to Reissner-Nordstr\"om black
hole. In the Reissner-
Nordstr\"om geometry the stability of Cauchy horizon is one of the most
interesting issues [3].
Recently we have shown that the Cauchy horizon is unstable because it is a
surface where
infalling tachyon got infinitely blueshifted [4].

In this paper, we study the  scattering of tachyon exterior of
charged 2D black hole. It is always possible to visualize the
black hole as presenting an effective potential barrier (or well) to the
on-coming waves.
One easy way of understanding the attributes of a physical system (black
hole) is to find
out how it reacts to external perturbations.
In the case of black hole, this is the very
popular because there is no other way in which an external
observer can explore the interior region of the horizon.
One may expect that some of
the incident wave will be irreversibly absorbed by the black hole, while
the remaining fraction will be scattered back to infinity.
There are two interesting quantum mechanical phenomena between
the tachyon  and  black hole.   One is the scattering of
tachyonic mode off  the black hole [5]. We calculate the reflection and
transmission coefficients
about this scattering.  The other is  the celebrated
Hawking radiation [6]. In the case of charged black holes, the potential
barrier  $V^{OUT}(r^*)$
turns out to be a complicated form.
Here the simplified form $V^{APP}(r^*)$ is used for scattering
analysis.  We derive the Hawking temperature of
the charged 2D black hole from the on-shell reflection coefficient by using the
Bogoliubov transformation.

{}From the conformal invariance of the heterotic 2D string theories, one can
derive
the $\beta$-function equations [2,7]

\begin{equation}
R_{\mu\nu} + 2 \nabla_\mu \nabla_\nu \Phi
-F_{\mu \rho} F_\nu^{~\rho} - {1\over 2} \nabla_\mu T \nabla_\nu T = 0,
\end{equation}

\begin{equation}
\nabla^2 \Phi - 2 (\nabla \Phi)^2 + {1 \over 2} \alpha^2 + {1 \over 4} F^2
+ {1 \over 2} T^2  = 0,
\end{equation}

\begin{equation}
 \nabla_\mu F^{\mu \nu} - 2 (\nabla_\mu \Phi) F^{\mu \nu} = 0,
\end{equation}

\begin{equation}
\nabla^2 T - 2 \nabla_\mu \Phi \nabla^\mu T + 2 T = 0,
\end{equation}
where $ F_{\mu \nu } = \partial_{[\mu} A_{\nu]}$ is the Maxwell field.
The above equations are also derived from the requirement that the fields
must be an extremum
of the low-energy string action [4,8]

\begin{equation}
S_{l-e} = \int d^2 x \sqrt{-G} e^{-2\Phi}
   \big \{ R + 4 (\nabla \Phi)^2 + \alpha^2 - {1 \over 2} F^2 - {1 \over 2}
(\nabla T)^2 +
 T^2 \big \}.
\end{equation}
After deriving the equations of motion, we take the  transformation for
convenience
\begin{equation}
-2\Phi \rightarrow \Phi,~~~ T \rightarrow \sqrt 2 T, ~~~-R  \rightarrow R.
\end{equation}
Then the equations of motion become
\begin{eqnarray}
&&R_{\mu\nu} + \nabla_\mu \nabla_\nu \Phi + F_{\mu\rho}F_{\nu}^{~\rho} +
\nabla_\mu T \nabla_\nu T = 0,  \\
&& (\nabla \Phi)^2 +  \nabla^2 \Phi - {1 \over 2} F^2  - 2 T^2 - 8 = 0,  \\
&&\nabla_\mu F^{\mu \nu} + (\nabla_\mu \Phi) F^{\mu \nu} = 0,   \\
&&\nabla^2 T + \nabla \Phi \nabla T + 2 T = 0,
\end{eqnarray}
where we set $\alpha^2 = 8$.
The charged black hole solution to the above equations are given by

\begin{equation}
\bar \Phi = 2 \sqrt 2 r,~~~ \bar F_{tr} = Q e^{-2 \sqrt 2 r},~~~ \bar T = 0,
{}~~~ \bar G_{\mu\nu} =
 \left(  \begin{array}{cc} - f & 0  \\
                             0 & f^{-1}   \end{array}   \right),
\end{equation}
with
\begin{equation}
f = 1 -  {m \over \sqrt 2}e^{- 2 \sqrt 2 r} + {Q^2 \over 8}e^{- 4 \sqrt 2 r},
\end{equation}
where $m$ and $Q$ are the mass and charge of the black hole, respectively.
For convenience, we take $m=\sqrt2$.  From $f=0$, the double horizons
($r_{\pm}$) are given by
\begin{equation}
r_{\pm} = { 1 \over 2 \sqrt 2 }\log \left[ {1 \pm  \sqrt { 1 - {Q^2 \over
2}} \over 2}\right],
\end{equation}
where $r_{+}(r_{-})$ correspond to the event (Cauchy) horizons.
To study the propagation of string fields, we introduce the perturbations
around
the background solution as [4]
\begin{eqnarray}
&&F_{tr} = \bar F_{tr} + {\cal F}_{tr} = \bar F_{tr} [1 - {{\cal F}(r,t)
\over Q}],        \\
&&\Phi = \bar \Phi + \phi(r,t),                       \\
&&G_{\mu\nu} = \bar G_{\mu\nu} + h_{\mu\nu}  = \bar G_{\mu\nu} [1 - h
(r,t)],     \\
&&T = \bar T + \tilde t \equiv \exp (-{\bar \Phi \over 2}) [ 0 + t (r,t) ].
\end{eqnarray}
Here we choose the metric perturbation ($h_{\mu\nu}$) in  such a way that
the background symmetry
should be restored in the perturbation level.
This is important in studying all black holes [9,10].
 One has to linearize (7)-(10) in order to obtain the equations governing
the perturbation.
Before we proceed,  we first check whether the graviton ($h$),  dilaton
($\phi$) and Maxwell
mode (${\cal F}$) are physically
 propagating modes in the 2D charged black hole background.
We consider the conventional counting of degrees of freedom.
The number of degrees of freedom for the gravitational field ($h_{\mu\nu}$) in
$d$-dimensions is $(1/2) d (d -3)$.  For $d=4$ Schwarzschild black hole,
we obtain two degrees of freedom. These correspond to Regge-Wheeler mode
for odd-parity perturbation
and Zerilli mode for even-parity perturbation.  We have $-1$ for $d=2$.
This means that in
two dimensions
the contribution of graviton is equal and opposite to that of a spinless
particle (dilaton).
 The graviton-dilaton modes are gauge degrees of freedom and thus cannot
appear in the
physical observables [9].
In addition, the Maxwell field has $d-2$ physical degrees of freedom.
For $d=2$, Maxwell field has no physical degrees of freedom.
Thus we insist that graviton-dilaton, and Maxwell modes
are gauge artefacts in the charged 2D black hole.
Since these are not physically propagating modes, it is not necessary to
study eqs.(7)-(9).
One remaining equation that describes a physically propagating mode  is
just the tachyon equation
(10), which can be rewritten as
\begin{equation}
f^2 \partial_r^2 t  + 2 \sqrt 2 f (1 - f) \partial_r t - 2 f (1 - f)t  -
\partial_t^2 t = 0.
\end{equation}
One  way to study the tachyonic equation  is to transform (18)
into the usual one-dimensional Schr\"odinger equation by eliminating the
linear derivative term.
Introducing the  coordinate transformation
$$r\to r^* \equiv g(r),$$
(18) can be rewritten as
\begin{equation}
f^2 g'^2 {\partial^2 \over \partial r^{*2}} t  + f \{ f g'' +  f' g'\}
{\partial \over \partial r^* }t - [\sqrt 2 ff' - 2 f (1 - f)]t
 - {\partial^2 \over \partial t^2} t = 0,
\end{equation}
where the prime ($\prime$) denotes the derivative with respect to $r$.
Requiring that the coefficient of the linear derivative vanish, one finds
the relation
\begin{equation}
g' =  {1 \over f}.
\end{equation}
Assuming $t( r^*,t ) \sim t_{\omega} ( r^* ) e^{-i\omega t}$,
one can cast (19) into  one-dimensional Schr\"odinger equation

\begin{equation}
\{ {d^2 \over dr^{*2}} + \omega^2 - V_T(r)\} t_{\omega} = 0,
\end{equation}
where the effective potential $V(r)$ is given by

\begin{equation}
V(r) = f(\sqrt 2 f' - 2  (1 - f)).
\end{equation}
When $Q=1$, $V(r)$ is a double-humped barrier well ($V^{IN}$) between
the Cauchy horizon and event horizon, while
it is just a potential barrier ($V^{OUT}$) outside the event horizon (see
Fig.1).
Here we are interested only in the scattering of tachyon by $V^{OUT}$
outside of the black hole.
When $Q$ is small, this is approximated  by the $Q=0$ case.
This case corresponds to the 2D dilaton black hole background as
\begin{equation}
\bar \Phi = 2 \sqrt 2 r,~~~ \bar F_{tr} = 0,~~~ \bar T = 0,
{}~~~ \bar G_{\mu\nu} =
 \left(  \begin{array}{cc} - \tilde f & 0  \\
                             0 & \tilde f^{-1}   \end{array}   \right)
\end{equation}
with
\begin{equation}
\tilde f = 1 -  e^{- 2 \sqrt 2 r}.
\end{equation}
The event horizon  is shifted from $r_{+} = 0$ (for $Q=0$) to $r_{+} = -
0.056$ (for $Q=1$) and
$r_{+} = - 0.0004$ (for $Q= 0.1$) . However, this shift does not affect the
scattering analysis outside
the  black hole. In this case we can introduce the explicit form of $r^*$

\begin{equation}
r^* \equiv g(r) = r + {1 \over 2 \sqrt 2} \ln (1 - e^{- 2 \sqrt 2 r}).
\end{equation}
Note that $r^*$ ranges from $- \infty$ to $+ \infty$, while $r$ ranges
from the event horizon of the black hole ($r_{+} = 0$) to $+ \infty$.
Using this coordinate, one can rewrite the exterior potential ($V^{OUT}$)
in terms of $r^*$ as

\begin{equation}
  V^{OUT}(r^*) = { 1 \over 2 (\cosh \sqrt{2} r^*)^2}\left[ 1 - {Q^2(6- 2
\exp(-2\sqrt 2 r^*))
\over 16(1+  \exp(2\sqrt 2 r^*))} -   { 3Q^4 \exp(-2\sqrt 2 r^*)
\over 64(1+  \exp(2\sqrt 2 r^*))^2} \right].
\end{equation}
Since the form of this potential is complicated, we will simplify the
potential for scattering
analysis. By noting that the second and last terms are small compared with
the first, one
approximate $V^{OUT}(r^*)$ as
\begin{equation}
  V^{APP}(r^*) = { 1 \over 2 (\cosh \sqrt{2} r^*)^2}\left[ 1 - \delta^2(Q)
\right]
\end{equation}
with $\delta^2(Q) = (\sqrt{4 \over 32}Q)^2$.
When $Q=0.1$, we have $V^{OUT}(r^*)=V^{APP}(r^*)$ for the whole regions
(see Fig.2).
Even for $Q=1$, we find that
$V^{OUT}(r^*) = V^{APP}(r^*)$ is a good approximation on the right branch
(see Fig.3).
Therefore, when the energy ($\omega^2$) is large and $Q\sim1$ or charge $Q$
is small $(<1)$,
 we expect that there is not much difference in scattering analysis
if we use $V^{APP}(r^*)$ instead of $V^{OUT}(r^*)$.

Let us thus study with the equation
\begin{equation}
\{ {d^2 \over dr^{*2}} + \omega^2 - V^{APP}(r^*)\} t_{\omega}(r^*) = 0.
\end{equation}
Following Ref. [11], the above is exactly solvable and $t_{\omega}(r^*)$ is
given by
\begin{eqnarray}
t_{\omega}(r^*) &=& C_1 (\cosh\sqrt{2} r^*)^{-2 \lambda}
 F( - \lambda + { i \omega \over 2 \sqrt{2} },
    - \lambda - { i \omega \over 2 \sqrt{2} }, {1 \over 2} ; z )
\nonumber \\
  &+& C_2 (\cosh\sqrt{2} r^*)^{- 2 \lambda} \sqrt{z}
 F( - \lambda + {1 \over 2} + { i \omega \over 2 \sqrt{2} },
    - \lambda + {1 \over 2} - { i \omega \over 2 \sqrt{2} }, {3 \over 2} ; z ),
\end{eqnarray}
where $z= -(\sinh \sqrt2 r^*)^2$ and $\lambda = ( \sqrt{4/32} Q - 1)/4$.
$C_1, C_2$ are two arbitrary constants and  can be determined by
imposing the appropriate boundary conditions.
The boundary conditions are as follows.   Asyptotically the solution
consists of both
ingoing ($\exp(- i\omega r^*): \longleftarrow$)
and outgoing ($ R \exp( i\omega r^*):\longrightarrow $), but at the horizon
it is purely ingoing
($ T \exp( - i\omega r^*):\longleftarrow $).
We call this type of solutions as $\{ t_{out} \}$ and
the corresponding vacuum state is defined as $|0 \rangle_{out}$. The
transmission amplitude $T$
 (coefficient ${\cal T} = |T|^2$) and reflection
amplitude $R$ (coefficient ${\cal R} = |R|^2$)  are given by

\begin{eqnarray}
T &=& \Big({1 \over 4}\Big)^{- i \omega/ \sqrt{2}} \exp(i \Delta)
{ \sinh (\pi \omega / \sqrt{2})  \over  \sinh (\pi \omega / \sqrt{2}) +
i\cosh(\pi\delta(Q)/2) },  \\
{\cal T} &=&
{ ( \sinh (\pi \omega / \sqrt{2}) )^2  \over ( \cosh(\pi\delta(Q)/2))^2 + (
\sinh (\pi \omega / \sqrt{2}) )^2 },  \\
R &=& \Big({1 \over 4}\Big)^{- i \omega/ \sqrt{2}} \exp(i \Delta)
{1 \over \cosh(\pi\delta(Q)/2) - i\sinh (\pi \omega / \sqrt{2}) },     \\
{\cal R} &=& {1 \over  (\cosh(\pi\delta(Q)/2))^2 + ( \sinh (\pi \omega /
\sqrt{2}) )^2 },
\end{eqnarray}
where the phase factor ($\exp(i\Delta)$) is given by
$$\exp(i \Delta) =
{ \Gamma ( i \omega / \sqrt{2} ) ( \Gamma ( 1/4 - i \omega / 2 \sqrt{2} ) )^2
\over \Gamma (- i \omega / \sqrt{2} ) ( \Gamma ( 1/4 + i \omega / 2
\sqrt{2} ) )^2 }. $$
As might be expected, one finds that ${\cal T} + {\cal R} = 1$.

Let us consider two limiting cases,
$\omega \rightarrow 0$ and $\infty$, for future references.
The first case  reduces to
${\cal T}(\omega \rightarrow 0) = 0$ and  ${\cal R}(\omega \rightarrow 0) = 1$.
This corresponds to the total reflection since the potential becomes
effectively an
infinite potential barrier.
The second case is given by
\begin{equation}
{\cal T} (\omega \rightarrow \infty) \approx 1, ~~~~~ {\cal R}(\omega
\rightarrow
 \infty ) \approx \exp(-  \sqrt 2 \pi \omega ) \equiv \exp(- {\omega \over
T_{st}}).
\end{equation}
For this case the energy $\omega$ of the mode is much larger than the height
($(1 - \delta^2 (Q) )/2$) of the potential barrier.
This corresponds to the classical picture of Boltzmann distribution with
the statistical
temperature $T_{st}= {1 \over \sqrt 2 \pi }$.

In order to obtain the Hawking temperature, let us introduce another
boundary conditions.
Asymptotically the wave is purely ingoing ($\longleftarrow$), but near the
event horizon
 it has both outgoing ($\longrightarrow$) and
ingoing parts ($ \longleftarrow$). We denote this type of solutions as $\{
t_{in}\}$ and
this vacumm state is defined as
$ |0 \rangle_{in}$.
 The vacuum states $|0 \rangle_{out}$ and $|0 \rangle_{in}$ form two
different bases
of which any state can be expanded.
Consequently these are two distinct Fock space vacuum states. The two sets
$\{ t_{out} \}$
and $\{ t_{in} \}$ are related by the Bogoliubov transformation [12,13].
It is then a standard calculation
to evaluate the expectaion value of the number opeartor $N^{out}$ in the
vacuum state
$ |0 \rangle_{in}$
\begin{equation}
_{in} \langle 0|N^{out}|0 \rangle_{in} ={1 \over |{C_2 \over C_1}|^2 -1 } =
  { {\cal R} \over 1- { \cal R} },
\end{equation}
where ${\cal R}$ is the  reflection coefficient for tachyon off the black
hole in (33).
Finally  we define the Hawking temperature as
\begin{equation}
_{in} \langle 0|N^{out}|0 \rangle_{in} = {1 \over { \exp({\omega \over
T^Q_{H}}) - 1}}.
\end{equation}
{}From (35) and (36), we read off the Hawking temperature as
\begin{equation}
T^Q_{H} = { \omega \over \ln \left[ 1+ ( {\sinh( {\pi \omega / \sqrt 2 })
\over \cosh(\pi \delta
(Q)/2)})^2 \right]}.
\end{equation}
We obtain in the limit of $\delta \to 0 (Q \to 0)$,
\begin{equation}
T^{Q\to 0}_{H} ={ \omega \over 2 \ln[ \cosh( \pi \omega / \sqrt 2 ) ]}.
\end{equation}
which is just the Hawking temperature of 2D dilaton black hole [14].
In the limit of $Q \to 0$ and $\omega \to \infty$, $T^Q_{H}$ reduces to the
statistical temperature
$T_{st}= {1 \over  \sqrt 2 \pi}$ in (34).
We see that for the $out$ observers the $in$ vacuum is full of particles in a
heat bath at the temperature $T^Q_{H}$.  Notice that this situation is
static (eternal black hole):
 although
the black hole emits the thermal radiation ($\longrightarrow$),
it does not lose its mass $m =\sqrt 2$.
This is because
it absorbs an equal amount of matter ($\longleftarrow$) from the heat bath
outside the horizon [5].

In conclusion, the charged 2D black hole is visualized as presenting an
potential barrier
$V^{OUT}(r^*)$ to on-coming tachyon wave in the exterior region of black hole.
Since this has a complicated form,
we approximate this into $V^{APP}(r^*)$. This is in a good agreement with
$V^{OUT}(r^*)$ for
either higher energy $\omega^2$ or small $Q$.   We use $V^{APP}(r^*)$
instead $V^{OUT}(r^*)$ for scattering
analysis. In this case one-dimensional Schr\"odinger equation is exactly
solvable.
We calculate the  reflection and transmission coefficients for scattering
of tachyon
off the charged 2D black hole.
The Hawking temperature is also derived from the  reflection coefficient by
Bogoliubov transformation.
In the limit of $Q \to 0$,
we recover the
Hawking temperature of the 2D dilaton black hole.

\acknowledgments

This work was supported in part by the Basic Science Research Institute
Program,
Ministry of Education, 1995, Project NO. BSRI-95-2413
and by NONDIRECTED RESEARCH FUND, Korea Research Foundation, 1994.

\figure{ Fig.1 : The $Q=1$ graph of the effective potential of tachyon  ($
V(r)$). This takes the
   double-humped barrier well ($V^{IN}$) inside the black hole, while it
takes a simple potential
 barrier ($V^{OUT}$) outside the black hole. The event
  horizon is at $r_{+}= -0.056$ and the Cauchy  horizon is at $r_{-}= -0.679$.}

\figure{ Fig.2 :The $Q=0.1$ graphs of potentials  outside black hole.
 The solid (dashed) lines refer $ V^{OUT}(r^*)(V^{APP}(r^*))$. In this case
one finds that
 $ V^{OUT}(r^*)=  V^{APP}(r^*)$ for whole regions.}

\figure{ Fig.3 :The $Q=1$ graphs of potentials  outside black hole.
 The solid (dashed) lines refer $ V^{OUT}(r^*) (V^{APP}(r^*))$. In this
case one finds that
 $ V^{OUT}(r^*) \sim V^{APP}(r^*)$ for right branch. }
\end{document}